\documentclass[11pt]{article}
\usepackage{moriond}

\newif\ifpdf %make a new conditional 
\ifx\pdfoutput\undefined %if no pdfout is being asked for \pdffalse
\else %else, we are being asked for pdfout,
\pdfoutput=1 %turn the flag ON, and 
\pdftrue %turn the conditional to TRUE. \fi
\fi

\ifpdf
\usepackage[pdftex]{color}
\usepackage[pdftex,bookmarks=true,bookmarksopen=true,bookmarksnumbered=true,a4paper=true,nesting=false,colorlinks=false]{hyperref}
\usepackage[pdftex]{graphicx}
\hypersetup{%
pdftitle = {B Physics at CDF},
pdfauthor = {Jonas Rademacker},
pdfsubject = {B Physics at CDF - Contribution to Moriond/EW 2004},
pdfkeywords = {CDF, CP violation, Bottom, Beauty, B meson, B hadron,
  CKM matrix, Pentaquark, Xi(1860), B Trigger, Two Track Trigger}
}
\else
\usepackage[dvips]{color}
\usepackage[dvips,bookmarks=true,bookmarksopen=true,bookmarksnumbered=true,a4paper=true,nesting=false,colorlinks=false]{hyperref}
\usepackage[dvips]{graphicx}
\fi

\usepackage{rotating}

\newcommand{\cdfNote}{0}
\newcommand{\cdfPreprint}{0}
\newcommand{\internal}{0}

\newcommand{\hepArchive}[1]{\href{http://arXiv.org/abs/#1}{[arXiv:{#1}]}}

\ifthenelse{\equal{\internal}{1}}{
 \newcommand{\citeCDF}[1]{~\cite{#1}}
 \newcommand{\bibCDF}[2]{\bibitem{#1} #2}
}{
 \newcommand{\citeCDF}[1]{}
 \newcommand{\bibCDF}[2]{}
}

\newcommand{\httpRefUscore}[2]{\href{#1}{\texttt{#2}}}

\def\be{\begin{equation}}
\def\ee{\end{equation}}
\def\bea{\begin{eqnarray}}
\def\eea{\end{eqnarray}}

\definecolor{darkgreen}{rgb}{0, 0.85, 0}
\definecolor{darkyellow}{rgb}{0.85, 0.85, 0}
\newcommand{\red}{}

\newcommand{\darkgreen}{}

\newcommand{\un}[2]{\ensuremath{\mathrm{#1 \, #2}}}

\newcommand{\units}[1]{\ensuremath{\mathrm{#1}}}

\newcommand{\degrees}[1]{\ensuremath{\mathrm{#1^{\circ}}}}

\newcommand{\E}[1]{\ensuremath{\cdot 10^{#1}}}

\newcommand{\prt}[1]{\ensuremath{{\rm #1}}}
\newcommand{\qrk}[1]{\ensuremath{#1}}

\newcommand{\Bso}{\prt{B_{s}^0}}

\newcommand{\Bdo}{\prt{B_{d}^0}}

\newcommand{\bbbar}{\qrk{b\overline{b}}}
\newcommand{\bbar}{\bbbar}

\newcommand{\order}[1]{\ensuremath{\mathcal{O}\!\left(#1\right)}}

\newcommand{\CP}{ \mbox{{\Large
      ${\not}\!$}\hspace{-0.3em} CP}
      }
\newcommand{\sCP}{ \mbox{{\large
      ${\not}\!$}\hspace{-0.3em} CP}
      }

\newcommand{\Gs}{\ensuremath{\Gamma_s}}
\newcommand{\DGs}{\ensuremath{\Delta\Gs}}

\newcommand{\fig}{.}
\newcommand{\talkFig}{.}
\newcommand{\talkNFig}{.}

\newcommand{\noteNumber}{CDF/PUB/BOTTOM/PUBLIC/7019}
\newcommand{\preprintNumber}{Conf-04/081-E}
\begin{document}
\ifthenelse{\equal{\cdfNote}{1}}{
\mbox{}\hfill\noteNumber}{}
\ifthenelse{\equal{\cdfPreprint}{1}}{
\mbox{}\hfill\preprintNumber}{}
\vspace*{4cm}
\title{B Physics at CDF}

\author{Jonas Rademacker\\ on behalf of the CDF Collaboration}

\address{Department of Physics, 1 Keble Road,\\
Oxford OX1 3RH, UK}

\maketitle\abstracts{
 Due to the large \bbar\ cross section at \un{1.96}{TeV}
 \prt{p-\bar{p}} collisions, the Tevatron is currently the most
 copious source of B hadrons. Recent detector upgrades for Run~II have
 made these more accessible, allowing for a wide range of B and \sCP\
 physics with B hadrons of all flavours. In this paper we present
 B-physics results, and, using the versatile hadronic Two Track
 Trigger, a search for \prt{\Xi(1860)}, from up to \un{240}{pb^{-1}}
 of data.\ifthenelse{\equal{\internal}{1}}{\it
 To appear in the proceedings of 39th Rencontres de Moriond on
 Electroweak Interactions and Unified Theories, La Thuile, Aosta
 Valley, Italy, 21-28 Mar 2004.
 This version differes from the version to be submitted to the
 Moriond/EW~2004 proceedings, by the extra links to the CDF
 internal notes describing the analyses.
                    }{}
 }

\section{Introduction}

 CDF has been taking data at Tevatron Run~IIa for about two years. For
 \prt{p\bar{p}} collisions at \un{1.96}{TeV}, the \bbbar\ production
 cross section is $\sigma_{\bbbar} \sim \un{0.1}{mb}$. 
 CDF has undergone major upgrades for Run~II, optimising its B~physics
 potential. The upgrades most relevant for CDF's B physics program
 include a new tracking system with a new, faster drift chamber, and
 new Silicon vertex trackers providing excellent proper time
 resolution, sufficient to resolve the expected fast oscillations in
 the \Bso\ system. The excellent impact parameter resolution is used
 for triggering on B-events. The muon coverage has been increased. A
 di-muon trigger efficiently finds \prt{B \to J/\psi X} decays.

 Here we present some of the wide range of analyses of the current CDF
 B physics program, which includes a wide range of studies, involving
 all types of B-hadrons, including leptonic as well as fully hadronic
 decays of \prt{B_d, B^{+}, B_s, B_c, \Lambda_b}.
 The impact-parameter based trigger also provides a very large sample
 of long-lived \prt{\Xi^-}. This has been used for a sensitive search for
 \prt{\Xi^{0}(1860) \to \Xi^- \pi^+} and \prt{\Xi^{--}\to \Xi^-\pi^-},
 which have been observed at NA49~\cite{Alt:2003vb} and are often
 interpreted as pentaquark states.

\section{Results from the Di-Muon Trigger}
\subsection[\qrk{b} Production Cross Section]{\qrk{b} Production Cross Section\citeCDF{CDFbProductionX}}
 The inclusive \qrk{b}-hadron production cross-section is measured
 from the b-fraction in the reconstructed \prt{J/\psi} sample up to
 February 2002 (\un{37}{pb^{-1}}). Combining this number with the
 inclusive \prt{J/\psi X} cross section, and the appropriate branching
 fractions, allows to calculate the absolute \qrk{b} production cross
 section. The long lifetime of B-hadrons is used to discriminate
 between prompt \prt{J/\psi} and \prt{J/\psi} from B-hadron
 decays. The total single b-quark cross-section integrated over one
 unit of rapidity is 
\[
\sigma(\prt{p\bar{p} \to \bar{b}
 X:\; |y| < 1.0}) = \un{29.4 \pm 0.6(stat) \pm 6.2(sys)}{\mu b}
\]
 where the largest contributions to the systematic error come from
 uncertainties in the acceptance and the inclusive B-hadron to
 \prt{J/\psi} branching ratio.

\subsection[Lifetimes]{Lifetimes\citeCDF{CDFLifetimesAndTransversity}\citeCDF{CDFLifetimesJoe}\citeCDF{CDFLambdaBLifetime}}
 Life time measurements in the heavy quark sector gain specific
 significance due to the precise predictions of Heavy Quark 
 Expansion~\cite{Uraltsev:1998bk}~\cite{Shifman:2000jv} thus
 providing a testing ground for this theoretical tool that is
 frequently used, for example to relate experimental measurements to
 CKM parameters like $\Gamma_d$ to $\left|V_{cb}\right|$ or $\Delta
 m_s/\Delta m_d$ to $\left|V_{ts}/V_{td}\right|$.

 Fully reconstructed hadronic \prt{B\to J/\psi X} decays, found with
 CDF's di-muon trigger, provide a clean method for measuring B lifetimes,
 free from the systematic uncertainties associated with semileptonic
 decays due to the missing momentum of the \prt{\nu}, and free from
 the lifetime bias in impact parameter-based trigger samples.
 Of specific interest at CDF are the lifetimes of the \prt{B_s} and
 \prt{\Lambda_b}, which are currently produced in large quantities
 only at the Tevatron.
\begin{table}
\caption[Lifetime Ratios]{Lifetimes and lifetime
 ratios in Run~II from
\prt{B_u^+ \to J/\psi(\mu^+\mu^-) K^{+}} (\un{240}{pb^{-1}}), 
\prt{B_d^0 \to J/\psi(\mu^+\mu^-)K^{(*)0}} (\un{240}{pb^{-1}}),
\prt{B_s^0 \to J/\psi(\mu^+\mu^-)\phi} (\un{240}{pb^{-1}}),
\prt{\Lambda_b \to J/\psi(\mu^+\mu^-)\Lambda} (\un{65}{pb^{-1}})
 compared with world average (HFAG~\cite{hfag}, results for PDG~04,
 and, for \prt{\Lambda_b}, results for PDG~02), Run~I
 results~\cite{RunILifetimes} and HQE
 predictions~\cite{CKMyellow2002}. Run~I results are from all channels
 combined, Run~II results from fully reconstructed \prt{
 J/\psi(\mu\mu) X} only.
\label{tab:CtauBJPsiRatios}
}
\begin{tabular}{cc}
\parbox{0.42\textwidth}{
 \includegraphics[width=0.43\textwidth]{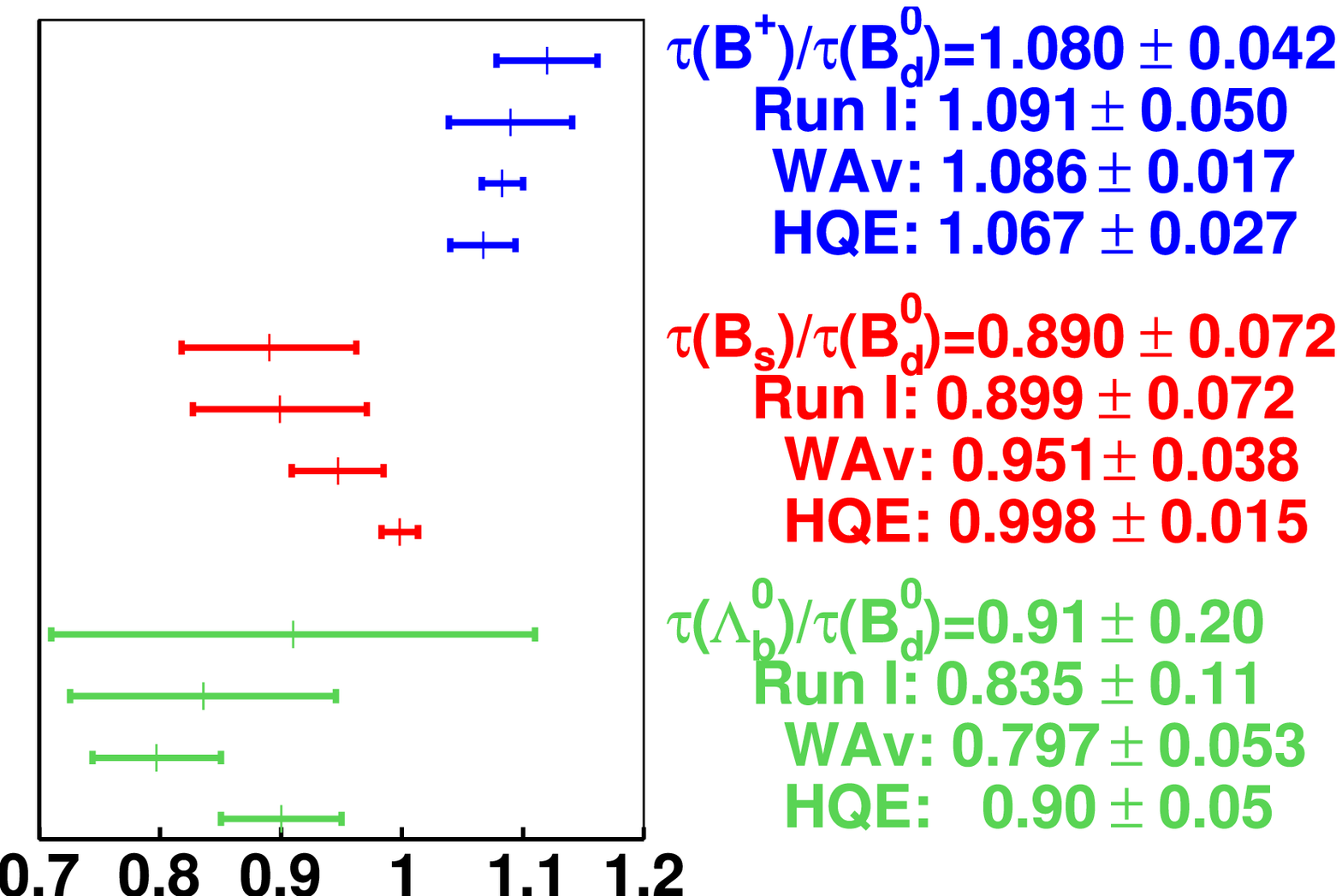}
}
&
\parbox{0.55\textwidth}{
\begin{tabular}{||ll||}
\hline\hline
 Channel & Result (\un{}{ps})
\\\hline
\prt{B_u^+ \to J/\psi(\mu^+\mu^-) K^{+}} & 
                    \un{1.662\pm 0.022 \pm 0.008}{}
\\\hline
\prt{B_d^0 \to J/\psi(\mu^+\mu^-)K^{(*)0}} & 
                    \un{1.539 \pm0.051\pm 0.008}{}
\\\hline
\prt{B_s^0 \to J/\psi(\mu^+\mu^-)\phi} &
                    \un{1.369\pm 0.100^{+0.008}_{-0.010}}{}
\\\hline
\prt{\Lambda_b^0 \to J/\psi(\mu^+\mu^-)\Lambda} &
                    \un{1.25\pm 0.25\pm 0.10}{}
\\\hline\hline
\end{tabular}
\vspace{0.5ex}\\
 Note that the Run~II result for \prt{B_s \to J/\psi \phi} is
 dominated by the (shorter) lifetime of the CP-even
 component.
}
\end{tabular}
\end{table}
 Lifetime results, and lifetime ratios, compared to theory
 predictions, Run~I results, and world averages, are summarised in
 Table~\ref{tab:CtauBJPsiRatios}.

\subsection[CP content of \prt{\it B_s \to J/\psi \phi}]{%
            CP content of \prt{\it B_s \to J/\psi \phi}\citeCDF{CDFPolJPsiPhi}}
% Ke Li, Colin Gay, Michael Schmidt (Yale)
\begin{table}
\begin{center}
\caption{Transversity-angle analysis in \prt{B_s \to J/\psi \phi} and
\prt{B_d \to J/\psi K^{*0}}. $A_0$ and
$A_{\parallel}$ are CP even decay amplitudes, $A_{\perp}$ is CP-odd,
normalised such that
\(
   \left|A_0\right|^2 
  + \left|A_{\parallel}\right|^2
  + \left|A_{\perp}\right|^2 
  \equiv 1.
\)
\label{tab:CPContentJPsiPhi}
}
\begin{tabular}{||c|c||}
\hline\hline
  \prt{B_s \to J/\psi \phi} & \prt{B_d \to J/\psi K^{*0}}\\
\hline
\parbox{0.465\textwidth}{\small
\setlength{\arraycolsep}{0.5ex}
\begin{eqnarray*}
 A_0            & = & 0.762 \pm 0.044 \pm 0.07 \\
 A_{\parallel}  & = & \left(0.433 \pm 0.199 \pm 0.011\right)
                      \; e^{i\left(2.08 \pm 0.51 \pm 0.06\right)} \\
 \left|A_{\perp}\right|      
                & = & 0.481 \pm 0.104 \pm 0.025
\end{eqnarray*}
}
&
\parbox{0.465\textwidth}{\small
\setlength{\arraycolsep}{0.5ex}
\begin{eqnarray*}
 A_0            & = & 0.796 \pm 0.022 \pm 0.012 \\
 A_{\parallel}  & = & \left(0.433 \pm 0.037 \pm 0.014\right)
                      \; e^{i\left(3.10 \pm 0.50 \pm 0.06\right)} \\
 A_{\perp}      & = & \left(0.422 \pm 0.050 \pm 0.027\right)
                      \; e^{i\left(0.18 \pm 0.26 \pm 0.02\right)}
\end{eqnarray*}
}
\\\hline\hline
\end{tabular}
\end{center}
\end{table}
 The measurement of the average lifetime in \prt{B_s \to J/\psi \phi}
 constitutes a first step towards a measurement of \DGs, the width
 difference between the long and short lived CP eigenstates, which has
 some sensitivity to new physics, especially when compared to the mass
 difference, $\Delta m_s$, which is also going to be measured at the
 Tevatron. The CP-even and odd contribution in \prt{B_s \to J/\psi
 \phi} can be disentangled by analysing the decay in terms of
 transversity angles, leading to the measurement of two CP even
 amplitudes $A_0$ and $A_{\parallel}$, and one CP-odd amplitude,
 $A_{\perp}$~\cite{Dighe:1995pd}. The CDF Run~II results for
 \un{192}{pb^{-1}} are shown in Table~\ref{tab:CPContentJPsiPhi}, for
 both \prt{B_s \to J/\psi \phi} and, as a cross check, \prt{B_d \to
 J/\psi K^{*0}}. The \prt{B_d} results are consistent with those from
 BaBar~\cite{Aubert:2001pe} and CLEO~\cite{Jessop:1997jk}. The phases of
 the amplitudes provide an interesting test of factorisation, which
 predicts the relative phases to be either $0$ or
 $\pi$~\cite{Yeh:1997rq}.  The amplitude measurements imply a CP-even
 content in \prt{B_s \to J/\psi \phi} of $77\% \pm 10\%$. Work is in
 progress to combine this technique with the lifetime analysis for a
 $\Delta\Gamma_s$ measurement.

\subsection[Search for New Physics with \prt{\it B_{d,s} \to \mu^+\mu^-}]{%
            Search for New Physics with \prt{\it B_{d,s} \to \mu^+\mu^-}\citeCDF{CDFMuMu}}
 While in the Standard Model, the branching ratio of \prt{B_{d,s} \to
 \mu^+\mu^-} is \order{10^{-9}}, which is below the sensitivity of the
 Tevatron, many New Physics models predict enhancements of this mode
 by several orders of magnitude, for example mSUGRA~\cite{Dedes:2001fv} 
 and SO(10) Symmetry Breaking models~\cite{Dermisek:2003vn}. In mSUGRA,
 the \prt{B_{d,s} \to \mu^+\mu^-} branching ratio is
 approximately~\cite{Dedes:2001fv}
\[
  \mathrm{BR_{mSUGRA}} \left(\prt{B_s \to \mu\mu}\right)
           \approx 10^{-6} \cdot {\red \tan^6\beta }
           \frac{M^2_{1/2}\units{GeV}^4}{(M^2_{1/2} + M^2_{0})^3}
\]
 which increases rapidly with large $\tan\beta$.

\begin{figure}
\caption{Search for \prt{B_{d,s} \to \mu^+ \mu^-} \label{fig:Bsmumu}}
{\small
 \begin{tabular}{*{3}{p{0.32\textwidth}}} 
 (a) Discriminating Variables:
 Mass, lifetime, $\Delta \phi$ and isolation ($p_t(\mu)$ divided
 by all $p_t$ in a cone around the $\mu$).
 & 
 (b) 1 event found in
 overlap of search windows - consistent with bgk estimate of
 $1.05\pm 0.30$ (\prt{B_d}), $1.07\pm 0.31$ (\prt{B_s}), $1.75\pm 0.34$
 (combined).
&
 (c) Projected and current sensitivity to \prt{B_s \to \mu\mu} at CDF,
 not including expected improvements due to increased \prt{\mu} coverage.
\\
\includegraphics[height=0.3\textwidth]{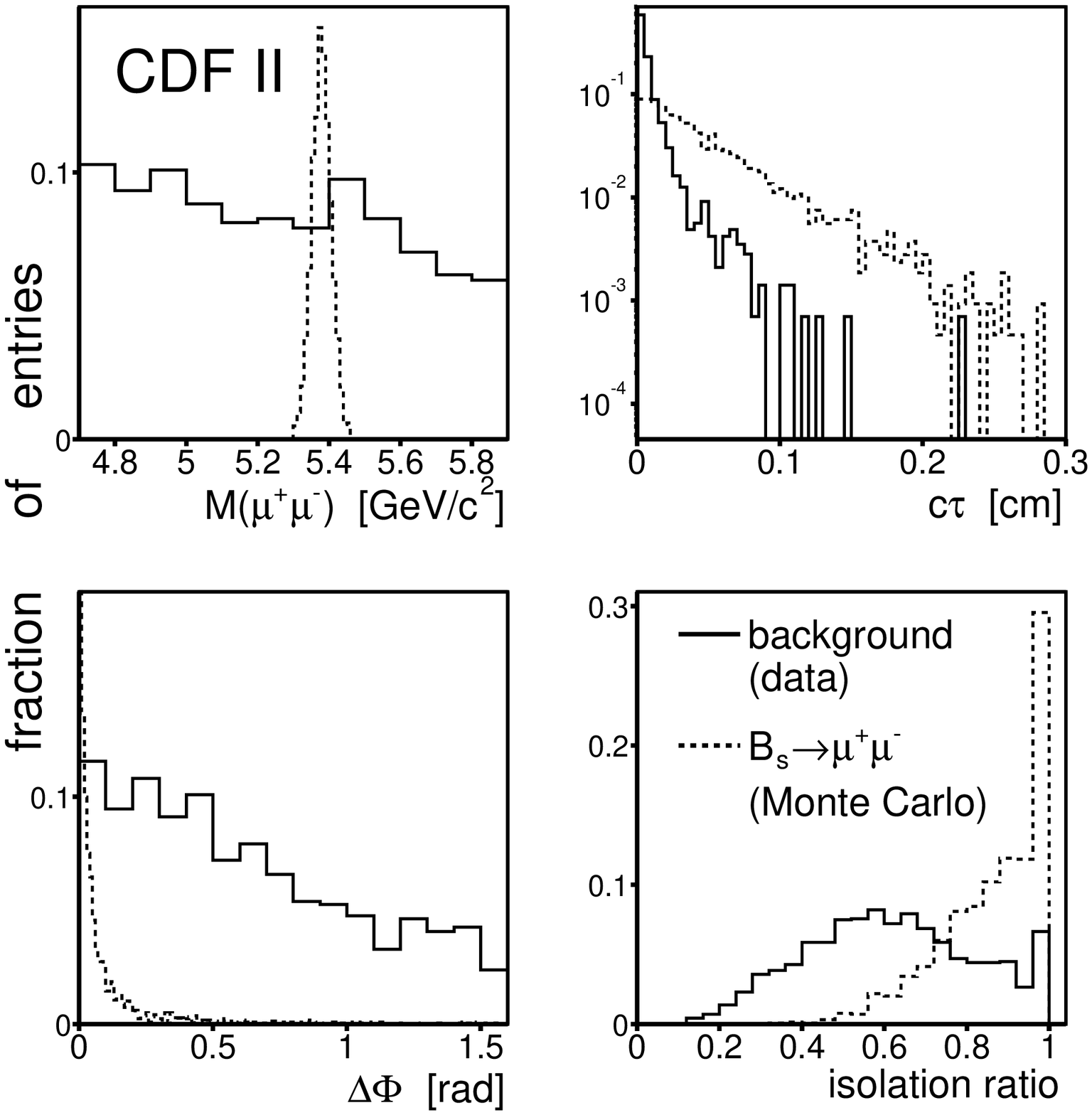}
&
\includegraphics[height=0.3\textwidth]{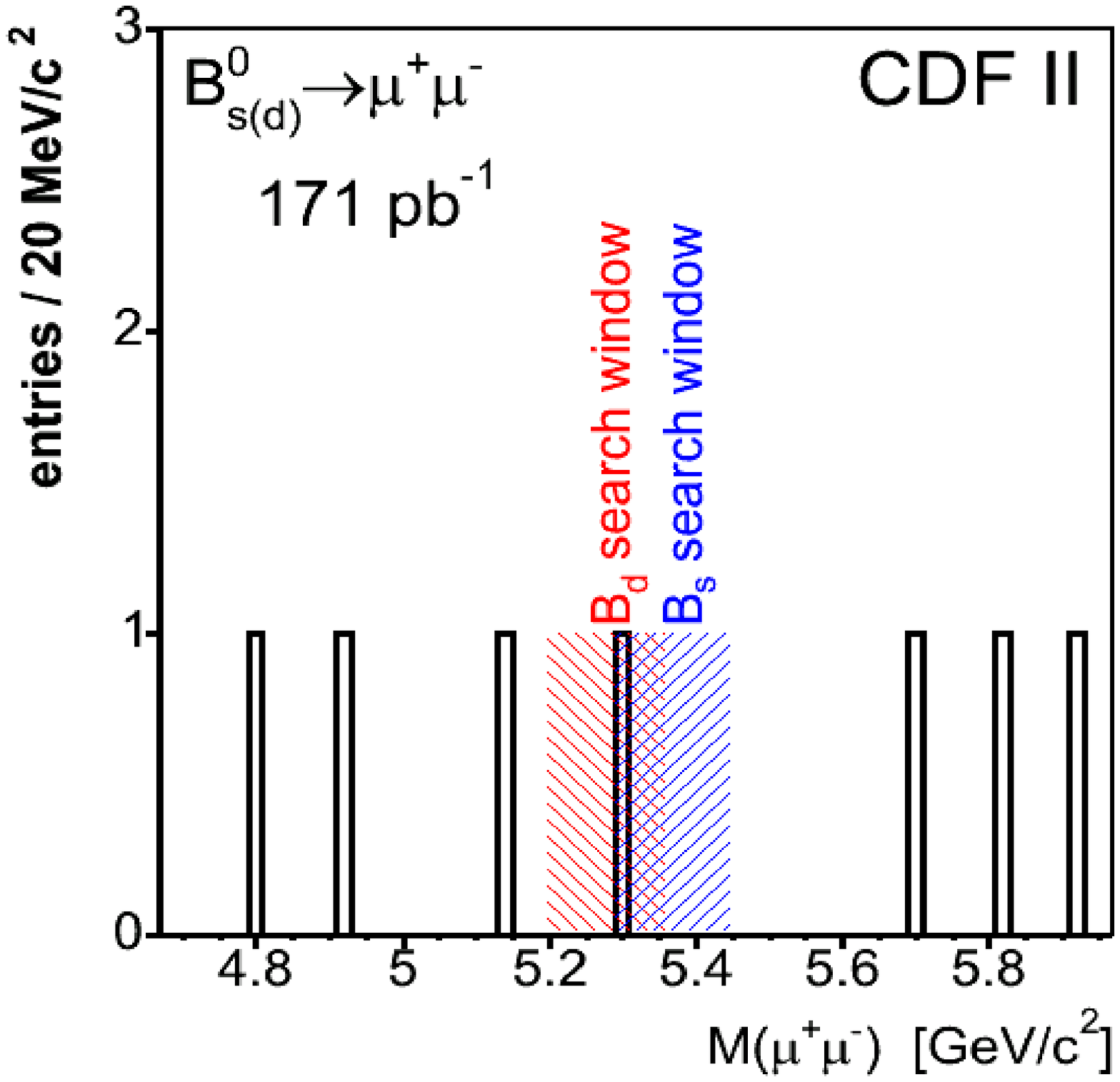}
&
\includegraphics[height=0.3\textwidth]{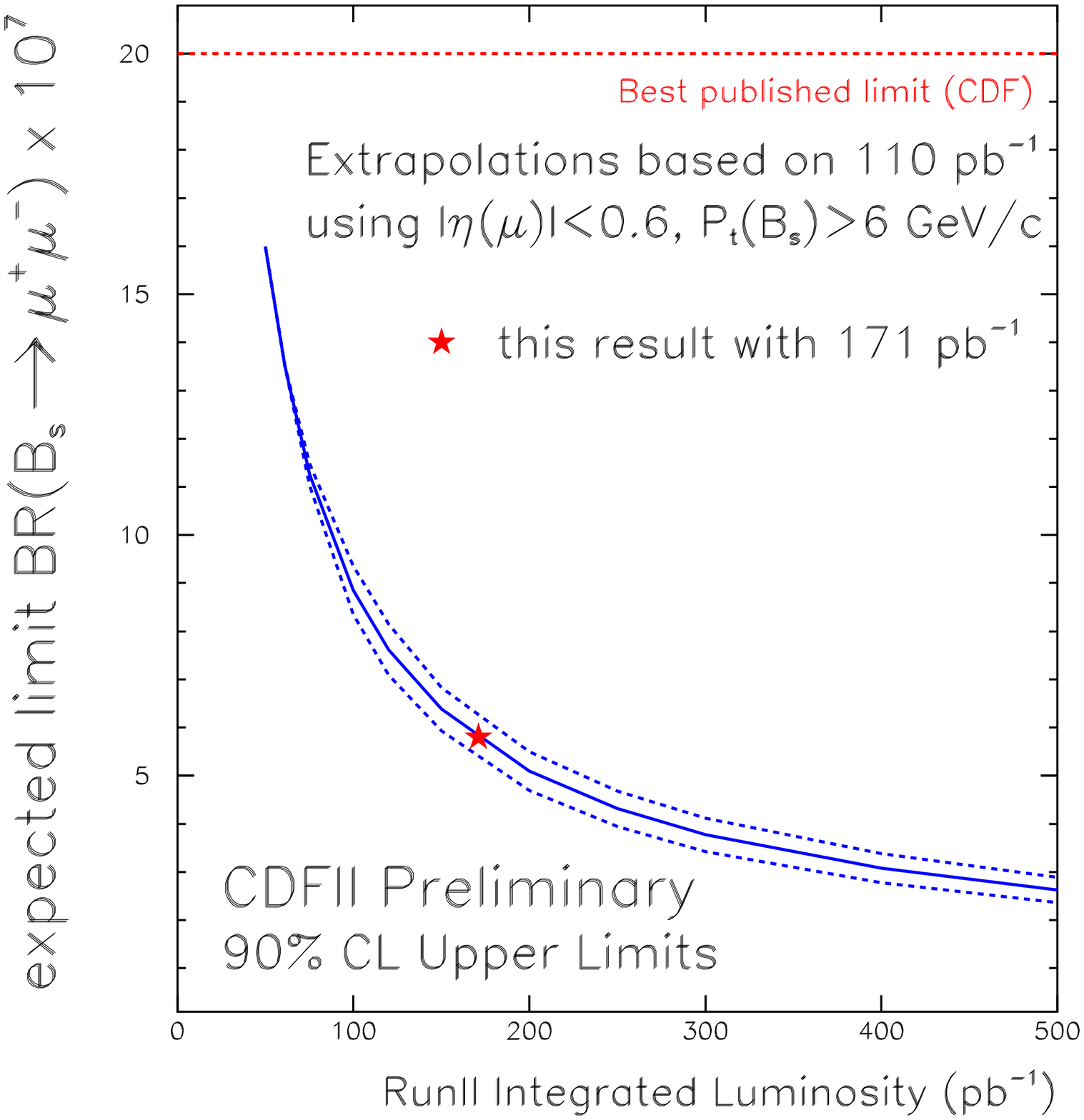}
\end{tabular}
}
\end{figure}
 The search for \prt{B_{d,s} \to \mu^+\mu^-} was performed as a blind
 analysis. The cuts were optimised using Monte-Carlo generated signal
 events and background events from real data. Signal and background
 distributions for the most important cuts are shown in Figure
 \ref{fig:Bsmumu} (a). After all cuts are applied, $1.05\pm 0.30$
 background events are expected in the \prt{B_d} mass window and
 $1.07\pm 0.31$ \prt{B_s} mass window, both are \un{200}{MeV} wide,
 and overlap.  The number of background events predicted for the
 combined mass window is $1.75\pm 0.34$. Several cross checks in real
 data have been performed before unblinding, for example using
 wrong-sign di-muon events (\prt{\mu^+\mu^+} and \prt{\mu^-\mu^-}),
 which yielded consistent results.  The total number of events found
 after unblinding is $1$ event in the overlap region of the two mass
 windows, as shown in Figure \ref{fig:Bsmumu} (b), resulting in the
 following $90\%$ confidence limits:
\vspace{0.5ex}\\\centerline{\fbox{
 BR(\prt{B_d \to \mu^+\mu^-})$ < 1.5\E{-7}$ (90\%CL)\hspace{2em}
 BR(\prt{B_s \to \mu^+\mu^-})$ < 5.8\E{-7}$ (90\%CL)
}}
\vspace{0.1ex}\\ which is, for the \prt{B_d}, similar to the results
 from BaBar and BELLE, and more than a factor of $3$ better than the
 previous best limit for \prt{B_s \to \mu\mu}, which was provided by
 CDF Run~I. The projected performance as a function of integrated
 luminosity, ignoring future improvements due to the expected increase
 in muon coverage, is shown in Figure~\ref{fig:Bsmumu}~(c).

\section{Results from the Impact Parameter-Based Hadronic B Trigger}
\subsection{CDF's Two Track Trigger}
 One of the most innovative improvements for B physics at CDF
 is the large-bandwidth hadron trigger, which triggers on the
 impact parameters of tracks at Level~2. 
\begin{figure}
\caption[The CDF hadron trigger]{The CDF hadronic 2-Track-Trigger. 
 $\Delta\phi$ is the angle between the tracks in the transverse
 plane. IP is the 2-D impact parameter of each of the two
 tracks. $\mathrm{L_{xy}}$ is the decay length in the transverse
 plane. The table on the left lists the trigger requirements. The
 figure on the right shows the IP resolution at trigger level.  } {
 \label{tab:hadronTriggerCDF}
\begin{tabular}{cc}
\fbox{\parbox{0.65\columnwidth}{
  {L1:} 2 XFT tracks, $p_t > \un{2}{GeV}$, $\Delta\phi < \degrees{135}$, $p_{t1} + p_{t2} > \un{5.5}{GeV}$.\vspace{1ex}\\
  {L2:}\\
  \begin{tabular}{p{0.3\columnwidth}|p{0.3\columnwidth}}
   2-body:\newline e.g. \prt{\Bdo \to \pi\pi} 
                      & Multi-body:\newline e.g. \prt{\Bso \to D_s \pi} 
   \\\hline
  $\un{100}{\mu m}<IP<\un{1}{mm}$ & $\un{120}{\mu m}<IP<\un{1}{mm}$        \\
   $\degrees{20} < \Delta\phi < \degrees{135}$ 
                      & $\degrees{2} < \Delta\phi < \degrees{90}$ \\
   $L_{xy}$\un{>200}{\mu m}
                      & $L_{xy}$\un{>200}{\mu m}         \\
   IP of B \un{<140}{\mu m} & --                            \\
  \end{tabular}\vspace{1ex}\\
  {L3:} Same with refined tracks \& mass cuts.
}}
&
\parbox{0.33\textwidth}{
\includegraphics[width=0.28\textwidth]{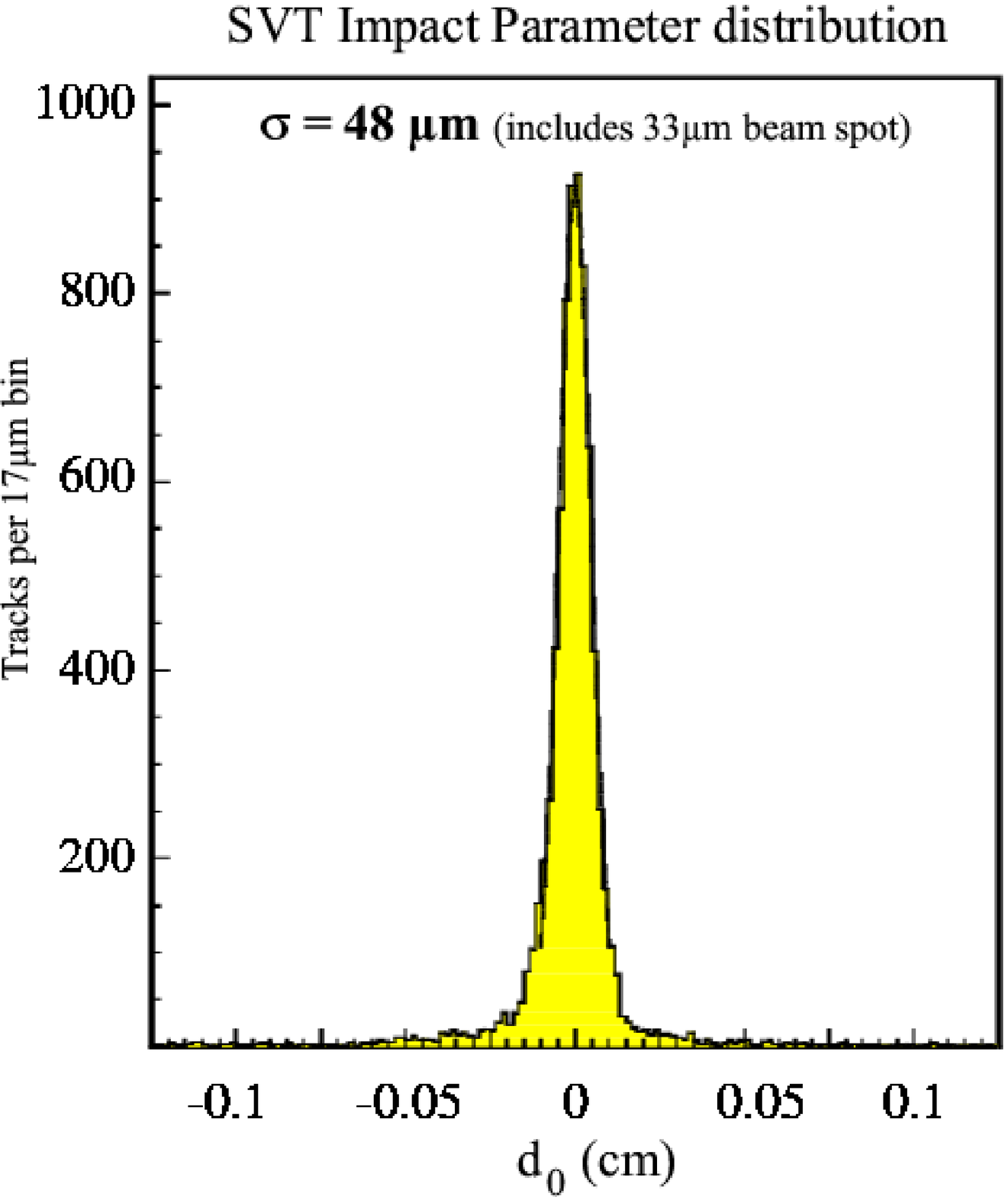}
}
\end{tabular}
}
\end{figure}
 The trigger requirements for the two scenarios, 2-body and multi-body
 B decays, are given in Figure~\ref{tab:hadronTriggerCDF}. 
 CDF's Two Track Trigger provides a unique sample of hadronic bottom
 and charm decays, that would otherwise be inaccessible, for example
 \prt{B^0\to\pi\pi} and \prt{B_s \to D_s \pi}.

\subsection[\prt{B\to hh}]{\prt{\it B\to hh}\citeCDF{CDFBhh}}
\begin{figure}
\caption{\prt{B \to hh}\label{fig:Bhh}}
{\small
\begin{tabular}{*{3}{c}}
\parbox{0.29\textwidth}{
 (a) 891 \prt{B \to hh} in \un{190}{pb^{-1}}}
&
\parbox{0.29\textwidth}{
 (b)
 \prt{K/\pi} sep. from 
\(
 \frac{\mathrm{d}E}{\mathrm{d}x} 
\)
}
& 
 (c)
 Kinematic variables (MC)
\\
\parbox{0.29\textwidth}{
\includegraphics[width=0.3\textwidth]{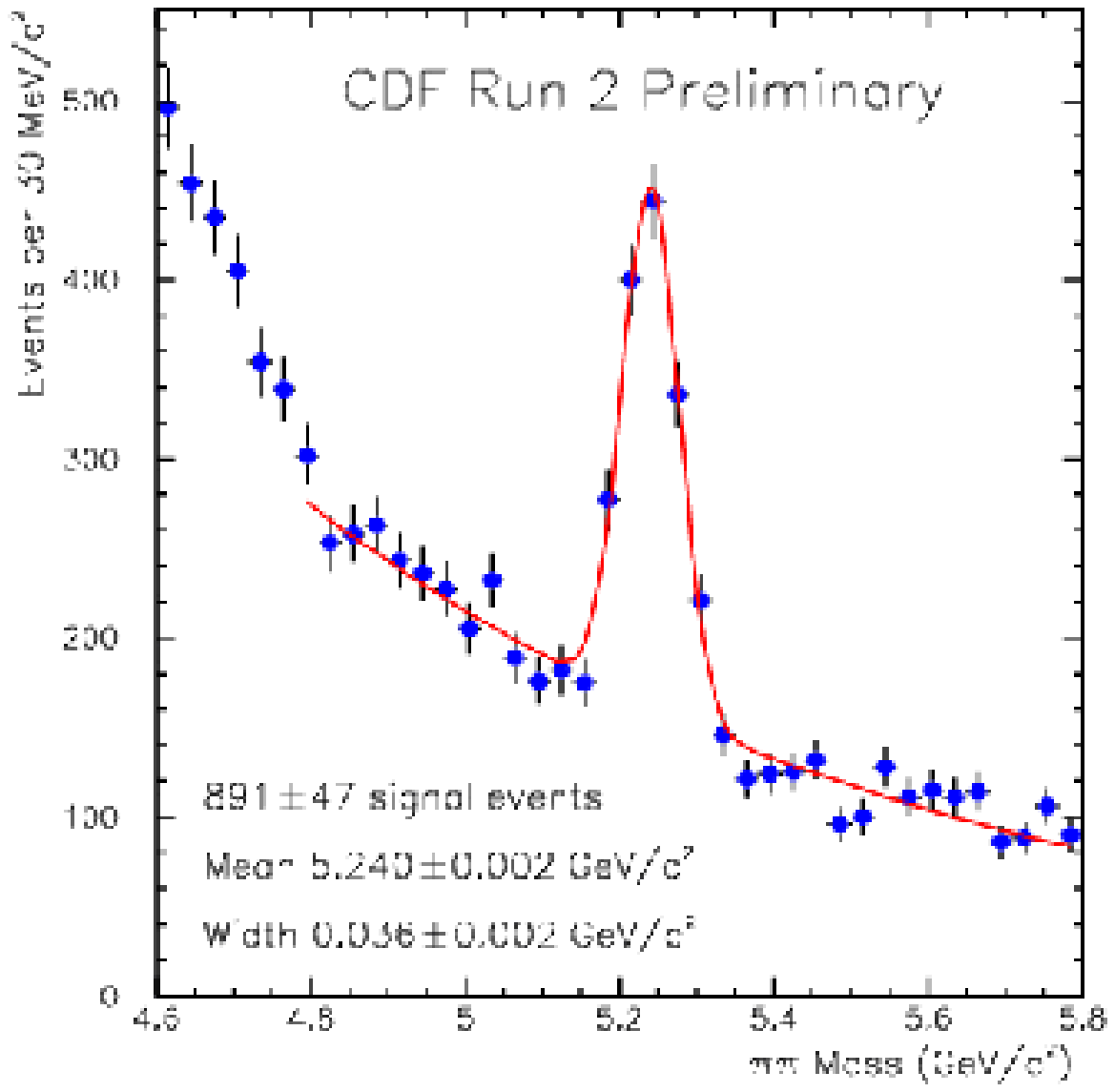}}
&
\parbox{0.29\textwidth}{
\includegraphics[width=0.3\textwidth]{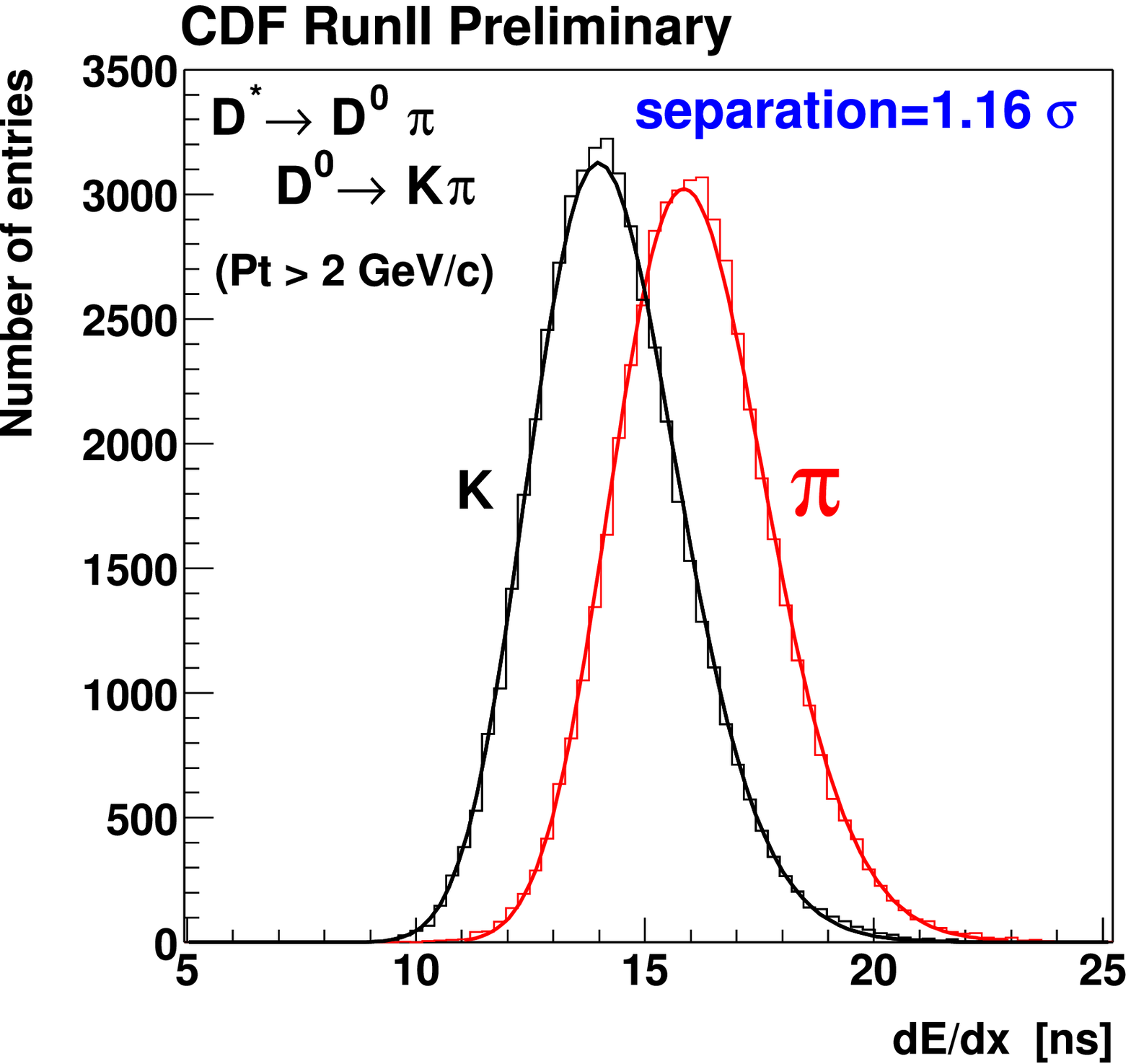}}
&
\parbox{0.39\textwidth}{
 \begin{tabular}{cc}
  \parbox{0.23\textwidth}{
  \includegraphics[width=0.28\textwidth]{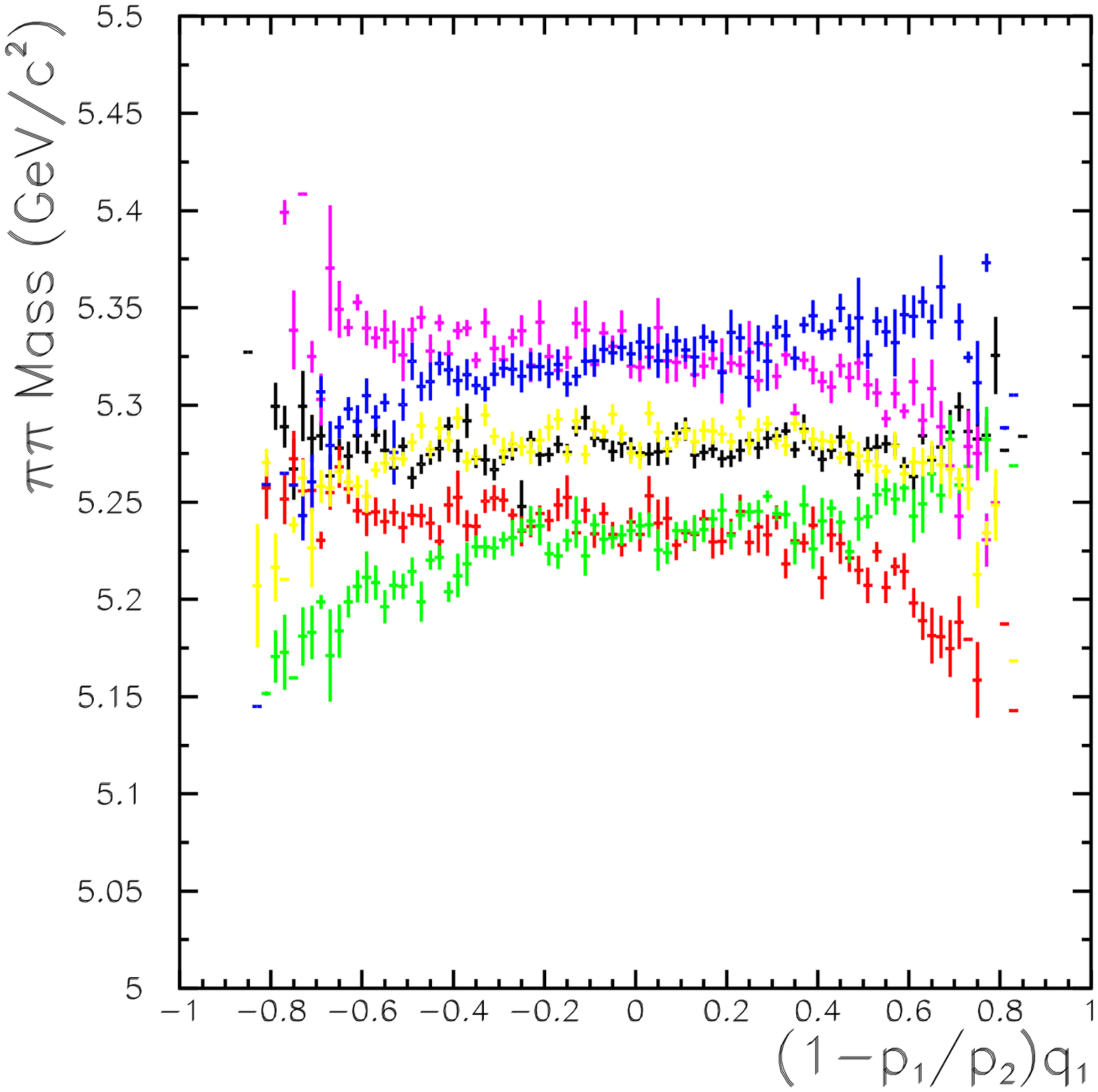}}
 &
  \parbox{0.15\textwidth}{\small
   \prt{\color{black} B_d \to \pi\pi}\\
   \prt{\color{darkyellow} B_s \to KK}\\
   \prt{\color{red} B_d \to K \pi}\\
   \prt{\color{darkgreen} \bar{B}_d \to K\pi}\\
   \prt{\color{blue}  B_s \to K\pi}\\
   \prt{\color{magenta} \bar{B}_s \to K\pi}
 }
 \end{tabular}
}
\end{tabular}
}
\end{figure}
 Figure \ref{fig:Bhh}~(a) shows the invariant mass of reconstructed
 \prt{B} to two-hadron events (assuming the hadrons are pions). About
 900 events are found. In order to discriminate the different decay
 modes, pions and kaons are separated using their specific energy
 loss, $\frac{\mathrm{d}E}{\mathrm{d}x}$. The \prt{\pi/K}
 discrimination using $\frac{\mathrm{d}E}{\mathrm{d}x}$ has been
 measured using \prt{D^{*}} decays and has been found to be
 $1.16\sigma$, as shown in figure \ref{fig:Bhh}~(b). Further
 discrimination between the different \prt{B\to hh} decay modes is
 achieved using decay kinematics, as shown in \ref{fig:Bhh}~(c).  The
 plot shows the reconstructed \prt{B} mass in Monte Carlo simulated
 \prt{B\to hh} events vs $(1-p_1/p_2)\cdot q_1$ for different decay
 modes. Here, $p_1$ is the smaller of the two momenta, $q_1$ is the
 charge of the particle with momentum $p_1$, and the mass is
 calculated assuming the decay products are pions. This led to the
 first observation of the decay \prt{B_s \to K^+K^-}. A summary of the
 results from analysing \prt{B\to hh} events in \un{65}{pb^{-1}} of
 data are given below:
\begin{itemize}
\setlength{\itemsep}{2ex plus2ex minus2ex}
 \item First observation of {\prt{B_s \to
       KK}: $90 \pm 24$}  out of $300$ \prt{B\to hh} events.
 \item Search for \CP\ in time-integrated rates\\
\(
   A_{CP}=\frac{   \Gamma(\prt{\red       \bar{B}^0_d \to K^- \pi^+} )
          - \Gamma(\prt{\darkgreen     {B}^0_d \to K^+ \pi^-} )
        }{  \Gamma(\prt{\red       \bar{B}^0_d \to K^- \pi^+} )
          + \Gamma(\prt{\darkgreen     {B}^0_d \to K^+ \pi^-} )
         }
         = { 0.02 \pm 0.15 \pm 0.017}
\)
\item Ratios of B.R.: \\
      \(  \frac{\Gamma(\prt{{B}^0_d \to \pi^+ \pi^-} )
                }{ \Gamma(\prt{{B}^0_d \to K^{\pm} \pi^{\mp}} )
                } = \un{0.26 \pm 0.11 \pm 0.06}{}
      \), 
      \(  \frac{\Gamma(\prt{{B}^0_s \to K^+ K^-} )
                }{ \Gamma(\prt{{B}^0_s \to K^{\pm} \pi^{\mp}} )
                } 
       = \un{2.71 \pm 0.73 \pm 0.35(f_s/f_d) \pm 0.81}{}
      \),\\
       where $(f_s/f_d)$ refers to the uncertainty due to the
      \prt{B_s}/\prt{B_d} production ratio.
\end{itemize}
 Results for \un{195}{pb^{-1}} should follow, soon. In the long term,
 these methods can be used to extract the CP-violating phase $\gamma$
 from a combined analysis of time-dependent decay rate asymmetries in
 \prt{B_d \to \pi\pi} and \prt{B_s \to KK}~\cite{Fleischer:1999pa}.

\subsection[\prt{D^0 \to hh}]{\it \prt{D^0 \to hh}\citeCDF{CDFDhh}}
 The Two Track Trigger also provides a huge charm signal, where the
 same methods can be applied. In the analysis presented here, only
 \prt{D^0} mesons from \prt{D^{*}} decays are used, which has two
 advantages: a very clean signal due to the highly effective cut on
 the difference between the reconstructed \prt{D^{*}} and \prt{D^0}
 mass, and the flavour of the \prt{D^0} is known from the charge of
 the \prt{D^{*}}. This allows a precise measurement of time-integrated CP
 asymmetries, which are expected to vanish in the Standard Model:
\begin{itemize}
\setlength{\itemsep}{2ex plus2ex minus2ex}
 \item
\(
   A_{CP\;KK}= \frac{   \Gamma(\prt{\red       \bar{D}^0 \to K^+ K-} )
          - \Gamma(\prt{\darkgreen     {D}^0 \to K^+ K^-} )
        }{  \Gamma(\prt{\red       \bar{D}^0 \to K^- K^+} )
          + \Gamma(\prt{\darkgreen     {D}^0 \to K^+ K^-} )
         }
         = {2.0\% \pm 1.2\% \pm 0.6 \%}
\)
 \item
\(
   A_{CP\;\pi\pi}= \frac{   \Gamma(\prt{\red       \bar{D}^0 \to \pi^+ \pi-} )
          - \Gamma(\prt{\darkgreen     {D}^0 \to \pi^+ \pi^-} )
        }{  \Gamma(\prt{\red       \bar{D}^0 \to \pi^- \pi^+} )
          + \Gamma(\prt{\darkgreen     {D}^0 \to \pi^+ \pi^-} )
         }
         = { 1.0\% \pm 1.2\% \pm 0.6 \%}
\)
\end{itemize}
 Branching ratios of \prt{D^0} mesons are also of some interest, for
example    
  \(  \frac{\Gamma(\prt{{D}^0 \to K^+ K^-} )
                }{ \Gamma(\prt{{D}^0 \to \pi^+\pi^-} )
                }
  \),
which is consistently larger experimentally, than theoretically
predicted. The following summarises the ratios of B.R. results:
\begin{itemize}
 \item
  \(  \frac{\Gamma(\prt{{D}^0 \to K^+ K^-} )
                }{ \Gamma(\prt{{D}^0 \to K^{\pm} \pi^{\mp}} )
                } = \un{9.96\% \pm 0.11\% \pm 0.12\%}{}
      \)
 \item
  \(  \frac{\Gamma(\prt{{D}^0 \to \pi^+ \pi^-} )
                }{ \Gamma(\prt{{D}^0 \to K^{\pm} \pi^{\mp}} )
                } = \un{3.608\% \pm 0.054\% \pm 0.12\%}{}
   \)
  \item
  \(  \frac{\Gamma(\prt{{D}^0 \to K^+ K^-} )
                }{ \Gamma(\prt{{D}^0 \to \pi^+\pi^-} )
                } = \un{2.762\% \pm 0.040\% \pm 0.034\%}{}
  \)
\end{itemize}

\subsection[\prt{B_s \to D_s \pi}]{\prt{\it B_s \to D_s \pi}\citeCDF{CDFBsDsPi}}
 The decay of \prt{B_s} to the flavour-eigenstate \prt{ D_s \pi} is
 the ``flagship mode'' for \prt{B_s} mixing at CDF. Being fully
 reconstructible (no missing \prt{\nu}), it provides for excellent
 time resolution - in topologically similar decays, CDF currently
 achieves \un{\sim 67}{fs}, and hopes to improve once the innermost Si
 layer has been fully commissioned and aligned. In \un{119}{pb^{-1}},
 $84\pm 11$~\prt{B_s \to D_s \pi} have been reconstructed with a
 signal to background ratio of $\sim 2$. The reconstruction efficiency
 has been increased since data taking has started and is now at $\sim
 1.6$ events per \un{}{pb^{-1}}. These data can be used to calculate
 the relative production$\times$B.R. in \prt{B_s \to D_s \pi} and
 \prt{B_d \to D \pi}:
\[
\frac{ f_s\cdot BR(\prt{\Bso \to D_s^-\pi^+}) 
     }{ f_d\cdot BR(\prt{\Bdo \to D^-\pi^+}) } = 0.35 \pm 0.05 \pm
     0.04 \pm 0.09 (BR)
\]
 where the last error is due to the uncertainty in the B.R. of the
 charm mesons.

\subsection[\prt{B_d} mixing]{\prt{\it B_d} mixing\citeCDF{CDFBMixing}}
 A further step towards measuring \prt{B_s} mixing is to make the
 somewhat easier measurement in the \prt{B_d} system and check for
 consistency with the well-established results from the B factories,
 and Run~1. About $1k$ \prt{B_d \to J/\psi K^*} and $5k$ \prt{B_d \to
 D \pi} events from \un{270}{pb^{-1}} were used for this
 measurement. The mass difference is extracted by measuring the
 oscillation frequency in time-dependent decay rate asymmetries. The
 asymmetries are between B decays that did not change flavour
 (e.g. \prt{B^0\to \bar{D}^0\pi^-}, neglecting Cabbibo suppressed
 decays)), and those that did (e.g. \prt{B^0\to D^0\pi^+}). In the
 measurement presented here, the flavour of the \prt{B^0} at birth was
 determined using same-side tagging only, which is based on the
 correlation of the \prt{B_d^0} or \prt{\bar{B}_d^0} flavour at birth,
 and the charge of the pion produced alongside, picking up the ``left
 over'' \qrk{\bar{d}} or \qrk{d} quark. (The same principle can be
 applied to \prt{B_s} mesons, using Kaon tags.) The tagging efficiency
 and dilution are measured using charged \prt{B} decays. The tagging
 power for same-side pion tagging is
\[
 \varepsilon D^2 = \left( 1.0 \pm 0.5 \pm 0.1 \right) \%.
\]
 where $\varepsilon = (63\pm 0.6)\%$ is the tagging efficiency
 (fraction of tagged events) and $D = (12.4\pm 3.3)\%$ the
 ``dilution'' defined as $D \equiv \left(1-2\omega\right)$, where
 $\omega$ is the mis-tag fraction. Note that a large ``dilution'',
 according to this definition, is a good thing. The tagging power
 $\varepsilon D^2$ describes the statistical power of the tag: $N$
 events before tagging are statistically equivalent $\varepsilon D^2
 \times N$ perfectly tagged events.
\begin{figure}
 \caption{Time-dependent decay rate asymmetries for \prt{B_d} mixing
 measurement, fitted simultaneously.\label{fig:BdMix}}
{\small
\begin{tabular}{cc}
   \prt{B_d \to J/\psi K^*} & \prt{B_d \to D \pi} 
\\
 \includegraphics[width=0.48\textwidth]{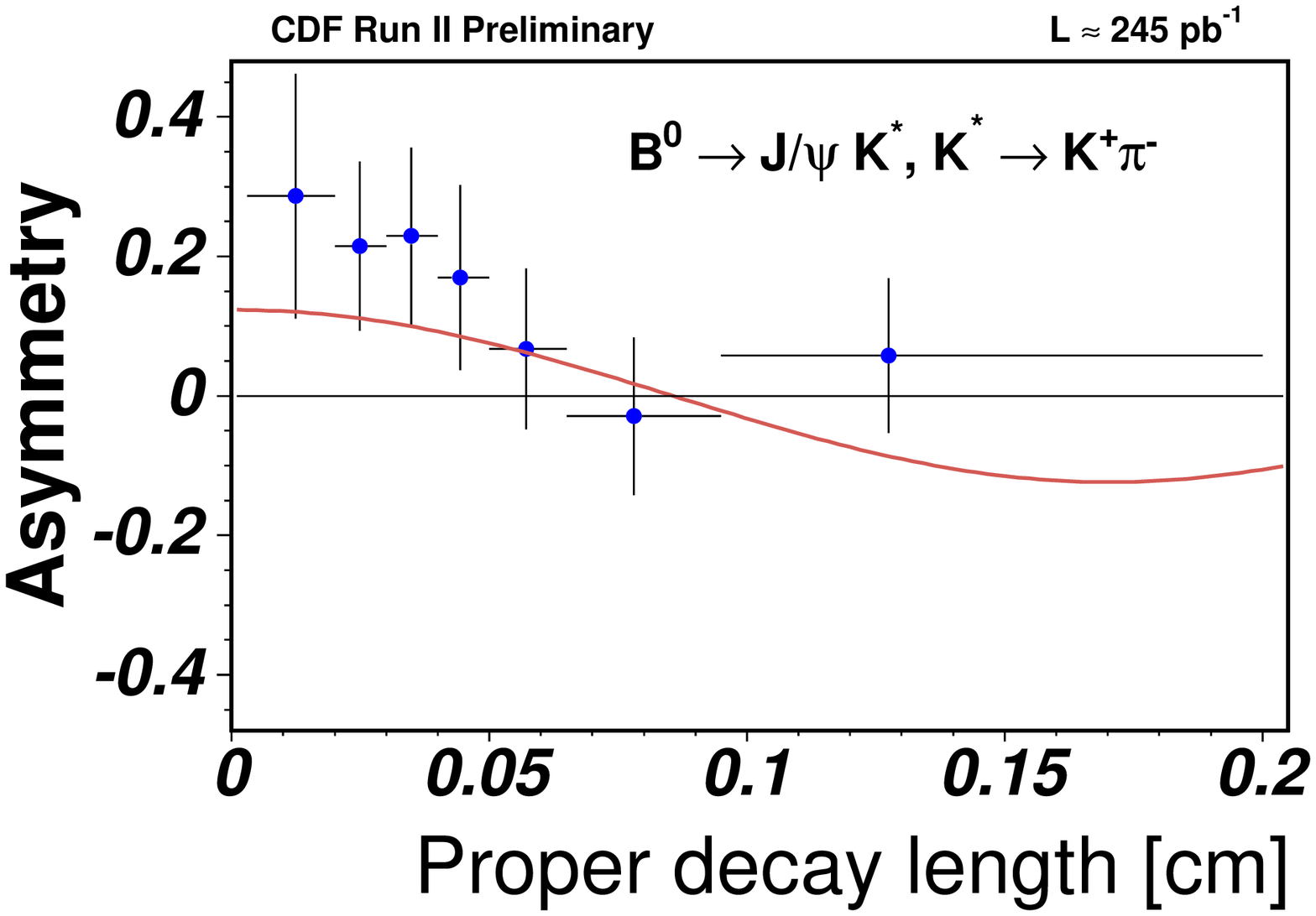}
&
 \includegraphics[width=0.48\textwidth]{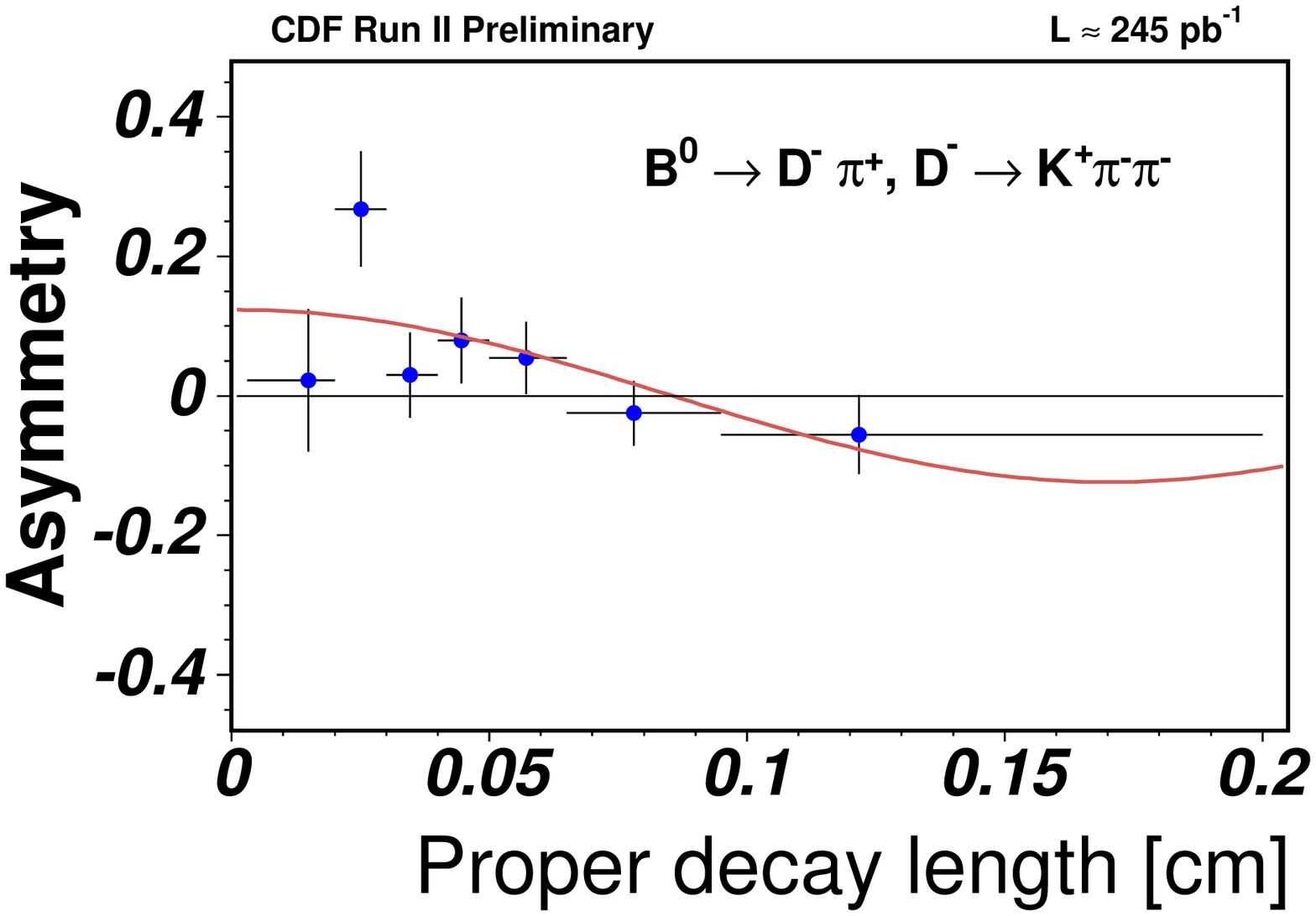}
\end{tabular}
}
\end{figure}
 A simultaneous fit to the time-dependent decay rate asymmetries in
 \prt{B_d \to J/\psi K} and \prt{B_d \to D \pi}, shown in Figure
 \ref{fig:BdMix} yields for the mass difference in the \prt{B_d}
 system:
\[
  \Delta m_d = \un{\left( 0.55 \pm  0.10 \pm 0.01 \right)}{ps^{-1}}
\]

\paragraph[Opposite side tagging]{Opposite side tagging\citeCDF{CDFMuonTag}\citeCDF{CDFJetQTag}}
 In independent studies, other tagging methods have been
investigated. Opposite side muon tagging yields a tagging power of
$\varepsilon D^2 = \left(0.660 \pm 0.093\right)\%$, jet charge tagging 
$\varepsilon D^2 = \left(0.419 \pm 0.024(stat)\right)\%$. Further
taggers are under investigation.

\subsection[Pentaquarks]{Pentaquarks\citeCDF{XiPiPenta}}
\label{sec:Pents}
 The impact-parameter based trigger does not only provide large
 numbers of bottom and charm mesons, but of all long lived particles,
 including the \prt{\Xi^-}. Combining this with a pion allows to
 search for the \prt{\Xi^{0}(1860)} and \prt{\Xi^{--}} observed at
 NA49~\cite{Alt:2003vb}, which is often interpreted as a pentaquark.

\begin{figure}
\caption{Searching for \prt{\Xi(1860)}.\label{fig:pents}}
{\small
\begin{tabular}{*{2}{p{0.46\textwidth}}}
 (a) \prt{\Xi^- \to \Lambda(p\pi) \pi^-} reconstruction.
&
 (b)  Invariant mass $M\left(\prt{\Xi^-},  \prt{\pi^{\pm}}\right)$. The peak at
 \un{1530}{MeV} is the well-known \prt{\Xi^0(1530)}.
\\
\includegraphics[width=0.45\textwidth]{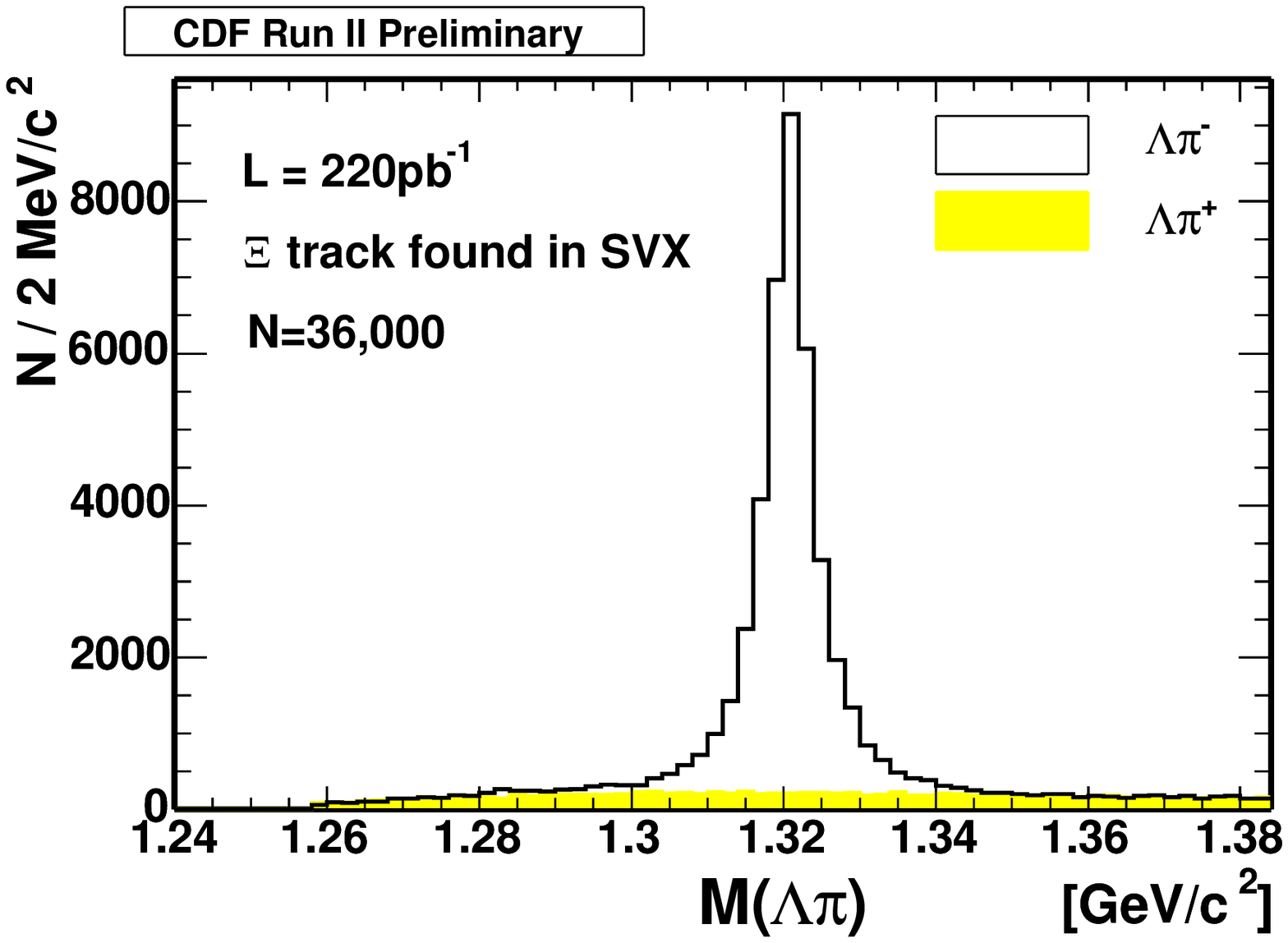}
&
\includegraphics[width=0.45\textwidth]{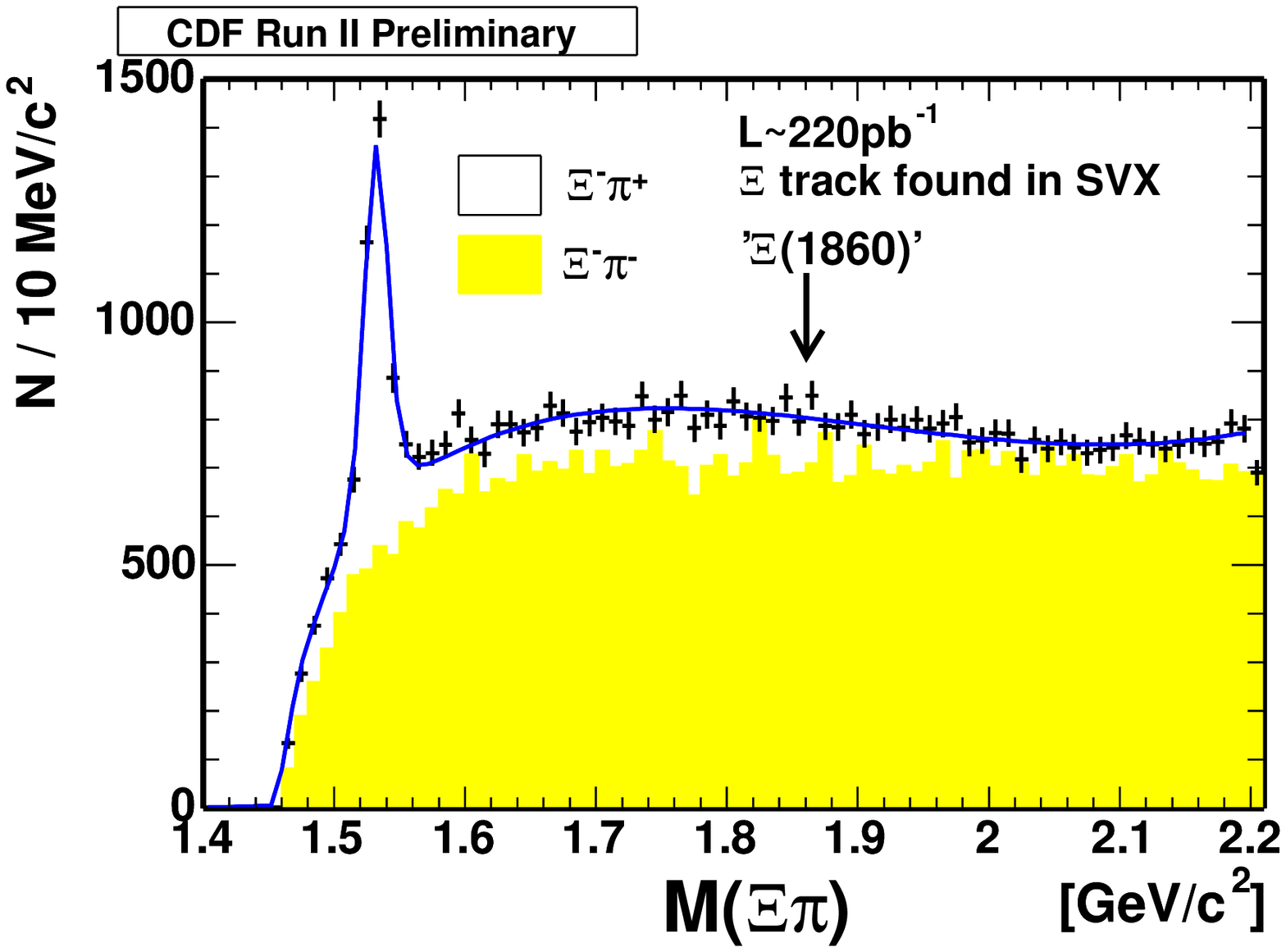}
\end{tabular}
}
\end{figure}
 CDF searches for the \prt{\Xi^{0}(1860)} and \prt{\Xi^{--}} in the
 decay modes \prt{\Xi^{0}(1860)\to \Xi^{-}\pi^+} and \prt{\Xi^{--}\to}
 \prt{\Xi^{-}\pi^-} with \prt{\Xi^- \to \Lambda(p\pi) \pi^-}.  The
 \prt{\Xi^-} lives long enough to leave hits in the Si detector before
 decaying. Requiring hits from the \prt{\Xi^-} in the Si provides a
 very efficient cut. Figure \ref{fig:pents} (a) shows the mass
 distribution a sample of $36,000$ \prt{\Xi^-}. The tiny background
 contribution, estimated from wrong-charge combinations, is
 superimposed as the shaded histogram.

 In a second step, the \prt{\Xi^-} is combined with a
 \prt{\pi^{\pm}}. Figure \ref{fig:pents} (b) shows the invariant mass
 distribution for same charge (shaded histogram) and opposite charge
 (black crosses) combinations of \prt{\Xi^-} and pions.  The line
 represents a fit to the opposite charge mass distribution. There is a
 clear peak at the well-known \prt{\Xi^0(1530)} resonance, that is
 used as a reference in this analysis.  However, neither the same sign
 nor the opposite sign combination show any evidence of a resonance at
 \un{1860}{MeV}. As a cross check, the analysis was repeated using the
 Jet20 trigger sample, that is not affected by an impact parameter
 cut. For $4k$ \prt{\Xi^-} in the Jet20 sample, no evidence of a
 \prt{\Xi(1860)} was found.  The $95\%$ upper confidence limits for
 the \emph{ratio} of \prt{\Xi(1860)} to the known \prt{\Xi^0(1530)}
 are:\\
\centerline{
\begin{tabular}{||c|c||}
\hline\hline
 \prt{\Xi^{-}\pi^+} (search) / \prt{\Xi(1530)}(control) & 0.07
\\\hline
 \prt{\Xi^{-}\pi^-} (search) / \prt{\Xi(1530)}(control) & 0.04
\\\hline\hline
\end{tabular}}

\section{Conclusion}
 Large numbers of B hadrons of all flavours are produced at the
 Tevatron. CDF has measured the $b$ production cross section in \prt{b
 \to J/\psi X} events. Fully reconstructed \prt{B\to J/\psi X} events
 have been used for precise lifetime measurements of \prt{B_d, B_s}
 and \prt{\Lambda_b} hadrons, which will provide a test of Heavy Quark
 Expansion. The CP content of \prt{B_s\to J/\psi\phi} has been
 measured using a transversity angle analysis, which will be combined
 with the lifetime measurment to extract $\Delta\Gamma_s$. Data from
 the leptonic B trigger were also used to obtain the best current
 limit on the B.R. of \prt{B_s\to\mu\mu}, one of the most sensitive
 probes of new physics at the Tevatron.

 CDF's high bandwidth Two Track Trigger provides a unique sample of
 hadronic B and Charm decays, including \prt{B\to hh}, which led to
 the first observation of \prt{B_s\to KK}, and will be used for CP
 violation studies as more data become available. First steps towards
 a \prt{B_s} mixing measurement have been taken with the reconstruction
 of \prt{B_s \to D_s \pi} events, and mixing measurements in the
 \prt{B_d} system.

 The huge sample of \prt{\Xi^-} found in the Two Track Trigger has
 been used for a sensitive search for \prt{\Xi(1860)}, which was not
 found. The B triggers will be used for many more pentaquark searches,
 especially those decaying to \prt{J/\psi} or \prt{D} and baryons.

\end{document}